\documentclass[letterpaper]{jpconf}

\usepackage{graphicx}% Include figure files
\usepackage{dcolumn}% Align table columns on decimal point
\usepackage{bm}% bold math
\usepackage{url}

\begin{document}

\title{Reactor monitoring and safeguards using antineutrino detectors}

\author{N~S~Bowden}
\address{Lawrence Livermore National Laboratory, 7000 East Ave., Livermore,
CA 94550, USA.}
 \ead{nbowden@llnl.gov}

\begin{abstract}
Nuclear reactors have served as the antineutrino source for many fundamental physics experiments. The techniques developed by these experiments make it possible to use these very weakly interacting particles for a practical purpose. The large flux of antineutrinos that leaves a reactor carries information about two quantities of interest for safeguards: the reactor power and fissile inventory.
Measurements made with antineutrino detectors could therefore offer an alternative means for verifying the power history and fissile inventory of a reactors, as part of International Atomic Energy Agency (IAEA) and other reactor safeguards regimes. Several efforts to develop this monitoring technique are underway across the globe.

\end{abstract}

\section{Introduction}
\label{sec:intro}

In the five decades since antineutrinos were first detected using a nuclear reactor as the source~\cite{reines}, these facilities have played host to a large number of neutrino physics experiments. During this time our understanding of neutrino physics and the technology used to detect antineutrinos have matured to the extent that it seems feasible to use these particles for nuclear reactor safeguards, as first proposed at this conference three decades ago~\cite{nu77}.

Safeguards agencies, such as the IAEA, use an ensemble of procedures and technologies to detect diversion of fissile materials from civil nuclear fuel cycle facilities into weapons programs. Nuclear reactors are the step in the fuel cycle at which plutonium is produced, so effective reactor safeguards are especially important. Current reactor safeguards practice is focused upon tracking fuel assemblies through item accountancy and surveillance, and does not include direct measurements of fissile inventory. While containment and surveillance practices are effective, they are also costly and time consuming for both the agency and the reactor operator. Therefore the prospect of using antineutrino detectors to non-intrusively \emph{measure} the operation of reactors and the evolution of their fuel is especially attractive.

The most likely scenario for antineutrino based cooperative monitoring (e.g. IAEA safeguards) will be the deployment of relatively small (cubic meter scale) detectors within a few tens of meters of a reactor core. Neutrino oscillation searches conducted at these distances at Rovno~\cite{rovno} and Bugey~\cite{bugey} in the 1990's were in many ways prototypes that demonstrated much of the physics required. Once the neutrino oscillation picture became clear at the start of this decade, all the pieces were in place to begin development of detectors specifically tailored to the needs of the safeguards community~\cite{firstpaper}. Longer range monitoring, e.g. that described in \cite{long_range}, would also be attactive, but will likely require significant advances before becoming feasible.
 %In this paper we will briefly describe the physics basis for this monitoring technique, the various efforts underway to develop it further, and the particular application in which the IAEA has expressed recent interest.

\section{Antineutrino Production in Reactors and Detection}

A more detailed treatment of this topic can be found in a recent review of reactor antineutrino experiments~\cite{reactor_nu_review}. Antineutrino emission by nuclear reactors arises from the beta decay of neutron-rich fragments produced in heavy element fissions. These reactor antineutrinos are typically detected via the inverse beta decay process on quasi-free protons in a hydrogenous medium (usually scintillator): $\bar{\nu}_e + p \rightarrow e^{+} + n$. Time correlated detection of both final state particles provides powerful background rejection. %Proposals to the use a different process (coherent neutrino-nucleus scattering) are noted below.

\begin{figure}[tb]
\begin{minipage}{3in}
\centering
\includegraphics*[width=2in]{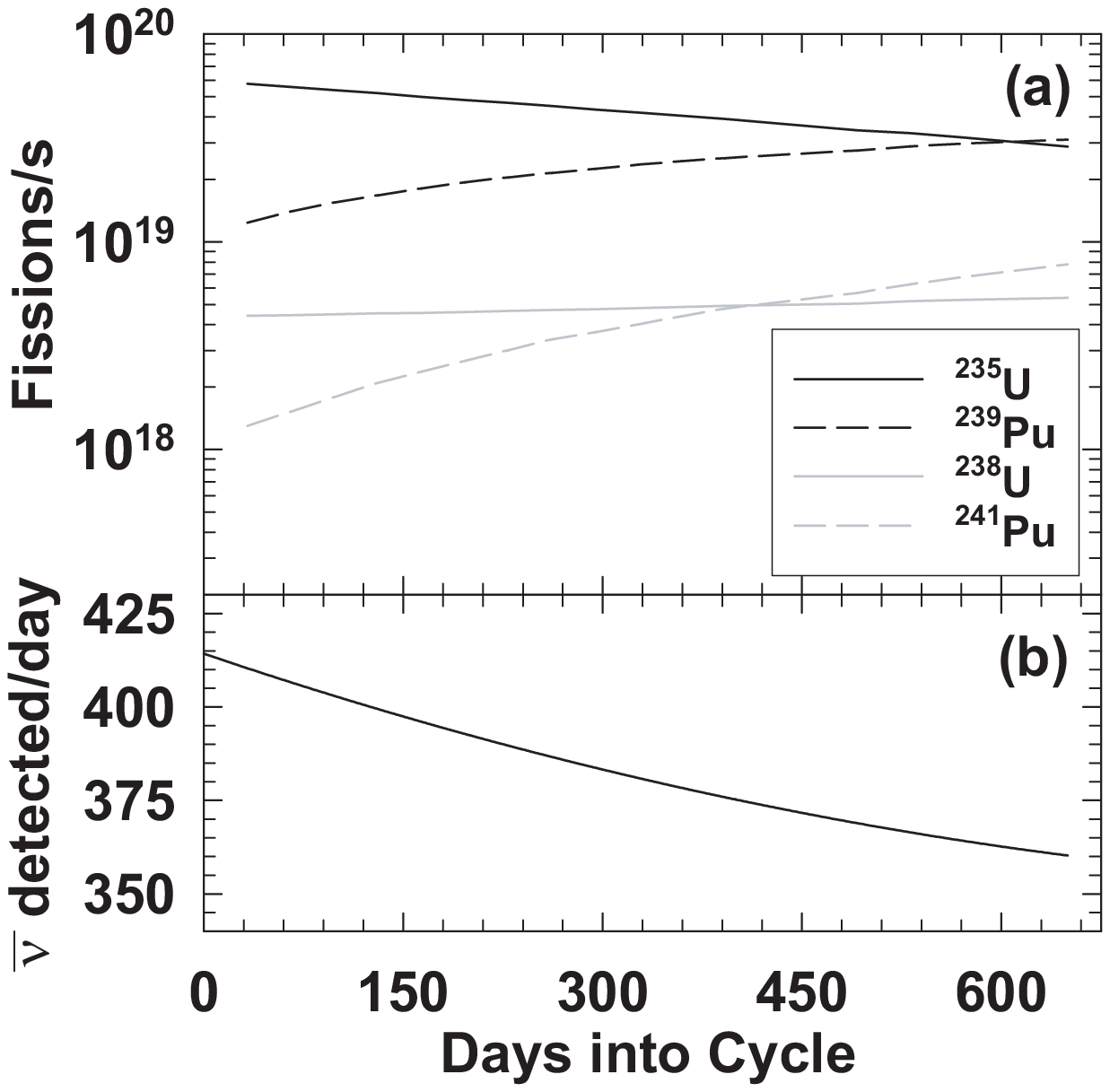}
\caption{The predicted (a) fission rates for the four main fissioning isotopes of a PWR
and (b) antineutrino detection rate in the LLNL/SNL SONGS1 detector throughout a reactor equilibrium fuel cycle.} \label{fig:fisrates}
\end{minipage}\hspace{2pc}%
\begin{minipage}{3in}
\centering\includegraphics*[width=2in]{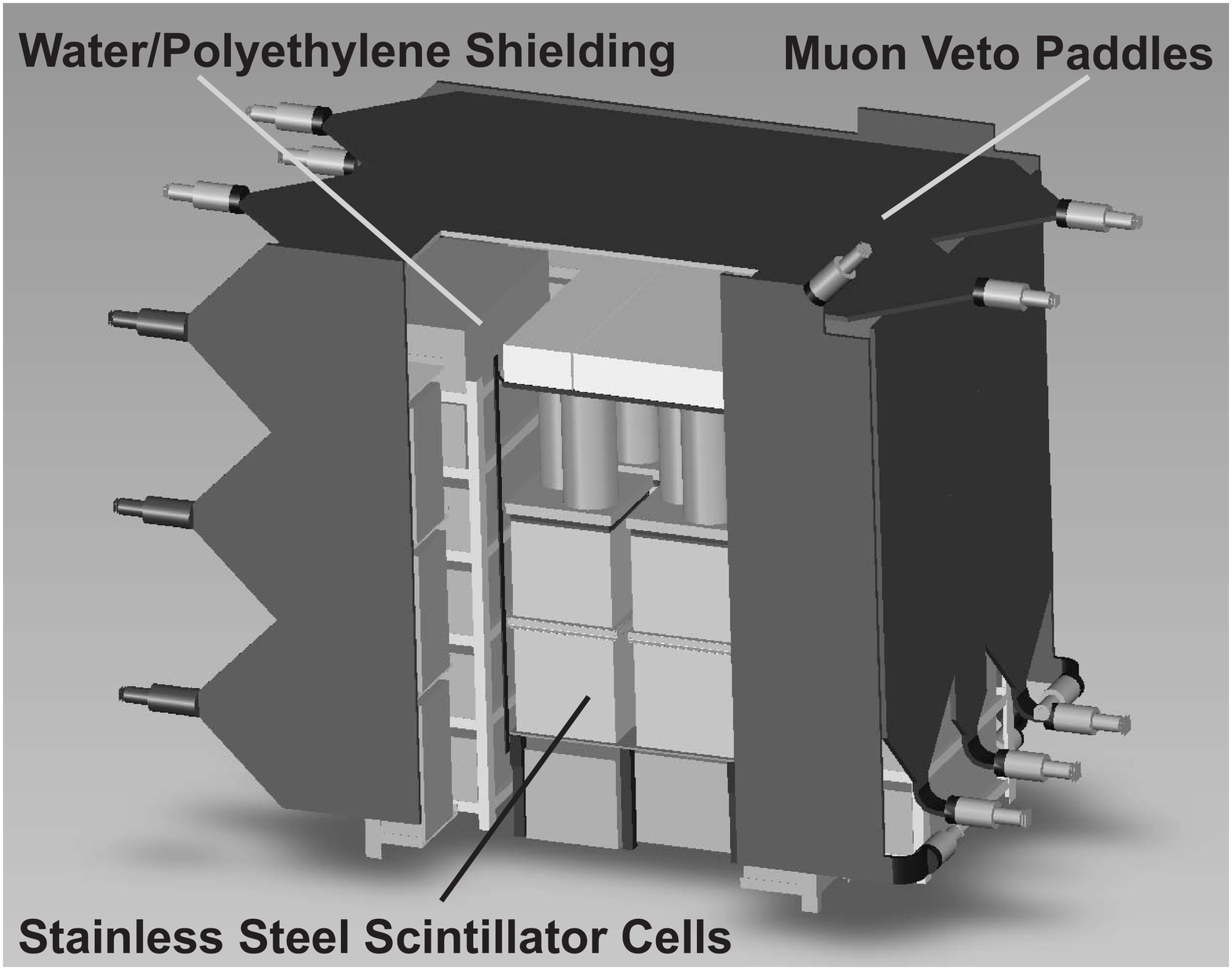}
\caption{A cut away diagram of the SONGS1 detector, showing the major
subsystems.} \label{fig:songs1}
\end{minipage}
\end{figure}

For the inverse beta process, the measured antineutrino energy spectrum, and thus the average number of detectable antineutrinos
produced per fission, differ significantly between the two major fissile elements, $^{235}$U and $^{239}$Pu
(1.92 and 1.45 average detectable antineutrinos per fission, respectively).
%The antineutrino energy spectrum, and thus the average number of detectable antineutrinos produced per fission, is significantly different for the two major fissile elements, $^{235}$U and $^{239}$Pu ($1.92$ and $1.45$ average detectable antineutrinos per fission, respectively).
Hence, as the reactor core evolves and the relative mass fractions and fission rates of $^{235}$U and $^{239}$Pu change (Fig.~\ref{fig:fisrates}a), the number of detected antineutrinos will also change. This relation between the mass fractions of fissile isotopes and the detectable antineutrino flux is known as the burnup effect.

Following the formulation of \cite{rovno}, it is instructive to write:
\begin{equation}
N_{\bar{\nu}}(t)  = \gamma \left(1+k(t)\right) P_{th}(t), \label{eq:nu_d_rate2}
\end{equation}
where $N_{\bar{\nu}}$ is the antineutrino detection rate, $P_{th}$ is the reactor thermal power, $\gamma$ is a constant encompassing all non varying terms (e.g. detector size, detector/core geometry), and $k(t)$ describes the change in the antineutrino flux due to changes in the reactor fuel composition.

Typically, commercial Pressurized Water Reactors (PWRs) are operated at constant thermal power, independent of the ratio of fission rates from each fissile isotope. PWRs operate for 1-2 years between refuelings, at which time about one third of the core is replaced. Between refuelings fissile $^{239}$Pu is produced by neutron capture on $^{238}$U. Operating in this mode, the factor $k$ -- and therefore the antineutrino detection rate $N_{\bar{\nu}}$-- decreases by about $0.1$ over the course of a reactor fuel cycle (Fig.~\ref{fig:fisrates}b), depending on the initial fuel loading and operating history.

Therefore, one can see from Eq.~\ref{eq:nu_d_rate2} that measurements of the antineutrino detection rate can provide information about both the thermal power of a reactor and the evolution of its fuel composition. These two parameters cannot, however, be determined independently by this method, e.g. to track fuel evolution one would need independent knowledge of the reactor power history. Measurement of the antineutrino energy spectrum may allow for the independent determination of the fuel evolution, and relative measurements over day to month time scales where $k$ varies little allow for tracking of short term changes in the thermal power. This last point may be of safeguards interest, since it is only when the reactor is off that the fissile fuel can be accessed at a PWR, and the integral of the power history constrains the amount of fissile Pu that can be produced.

\section{Global efforts to develop safeguards antineutrino detectors}

There are many efforts underway around the world to explore the potential of antineutrino based reactor safegaurds. The evolution of these efforts is summarized in the agenda of the now regular Applied Antineutrino Physics (AAP) Workshops~\cite{aap0,aap1,aap2,aap3}. At present, these efforts are funded by a variety of national agencies acting independently, but there is frequent communication between the physicists involved at the AAP meetings. This nascent AAP community is hopeful that recent IAEA interest (Sec.~\ref{sec:iaea}) will result in a formal request from the IAEA to the National Support Programs (the programs within member states that conduct research and development requested by the IAEA). Such a request that clearly laid out the the needs of the agency with respect to detector size, sensitivity, etc, would allow for better coordination between these respective national programs, and would be an important step towards the development of device for use under IAEA auspices.

\subsection{Effort in Russia}

As mentioned above, the concept of using antineutrinos to monitor reactor was first proposed by Mikaelian, and the Rovno experiment~\cite{rovno} was amongst the first to demonstrate the correlation between the reactor antineutrino flux, thermal power, and fuel burnup. Several members of the original Rovno group continue to develop antineutrino detection technology, e.g. developing new Gd liquid scintillator using a Linear Alkylbenzene (LAB) solvent. They propose to build a cubic meter scale detector specifically for reactor safeguards~\cite{russia}, and to deploy it at a reactor in Russia.

\subsection{Effort in the U.S.A.}

A collaboration between the Sandia National Laboratories (SNL) and the Lawrence Livermore National Laboratory (LLNL) has been developing antineutrino detectors for reactor safeguards since about 2000. Our particular focus is on demonstrating to both the physics and safeguards communities that antineutrino based monitoring is feasible. This involves developing detectors that are simple to construct, operate, and maintain, and that are sufficiently robust and utilize materials suitable for a commercial reactor environment, all while maintaining a useful sensitivity to reactor operating parameters.

\subsubsection{The SONGS1 detector}

The SONGS1 detector~\cite{secondpaper} was operated at the San Onofre Nuclear Generating Station (SONGS) between 2003 and 2006. %Its design was informed by all of the requirements listed in the preceding paragraph.
The active volume comprised $0.64$ tons of Gd doped liquid scintillator contained in stainless steel cells (stainless was used to avoid any chance of acrylic degradation and liquid leakage). This was surrounded by a water/polyethylene neutron-gamma shield and plastic scintillator muon veto (Fig.~\ref{fig:songs1}).
%The entire device was approximately cubic with each side measuring about $2.5$~m .

The detector was located in the tendon gallery of one of the two PWRs at SONGS, about $25$~m from the reactor core and under about $30$~m.w.e. overburden. Galleries of this type, which are part of a system for post-tensioning the containment concrete, are a feature of many, but not all, reactor designs. It may therefore be important to consider detector designs that can operate with little or no overburden.
%The compact size of the active volume means that a large fraction of the gamma rays released when a neutron captures on Gd are lost. This is a significant factor in the relatively low detection efficiency of $\approx 10\%$. Nonetheless, given the high antineutrino flux in the tendon gallery ($\approx 10^{17}$/(m$^2$s)), around $400$ reactor antineutrinos were detected per day.

\begin{figure}[tb]
\begin{minipage}{3in}
\centering
\includegraphics*[width=2.5in]{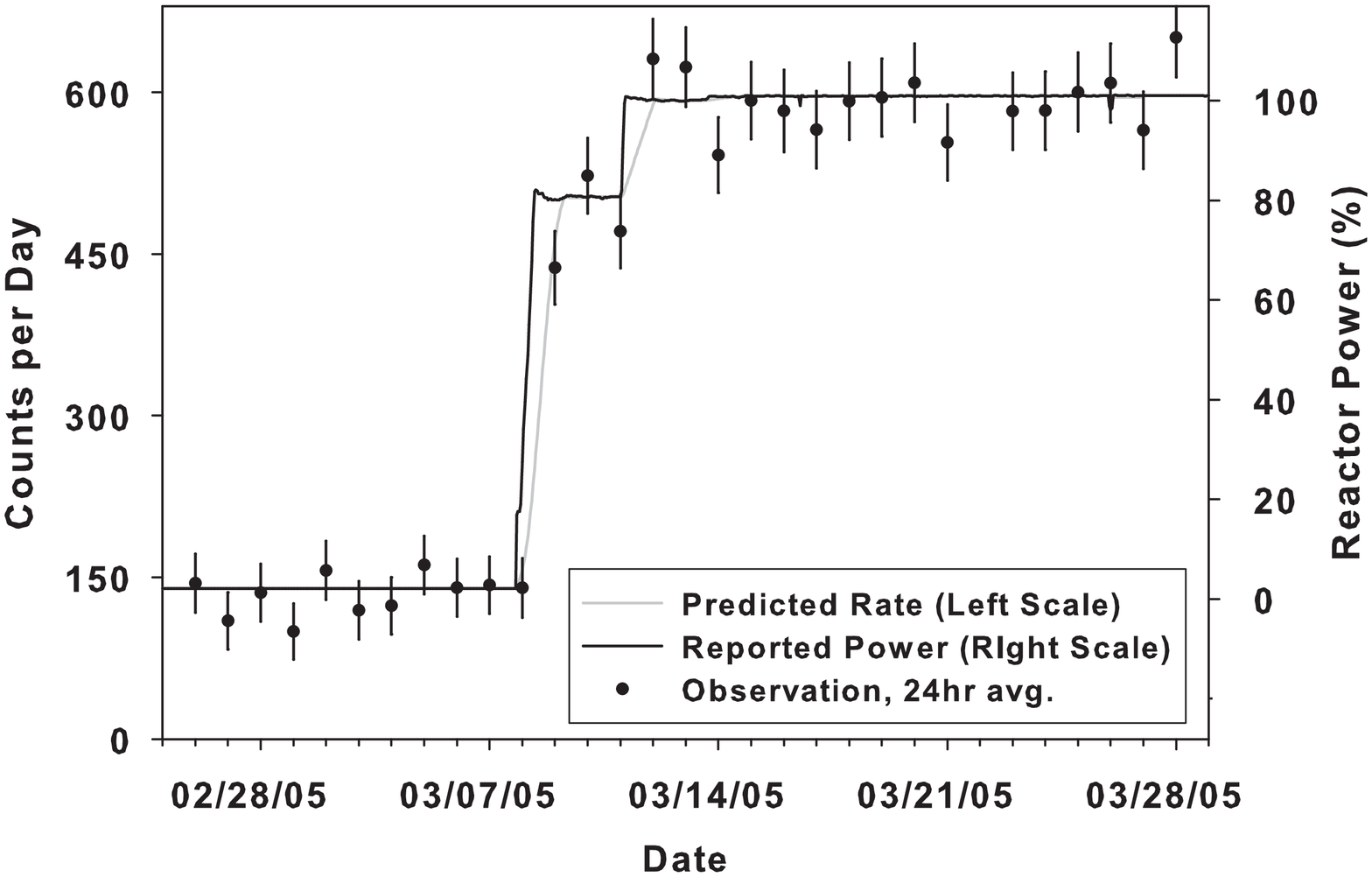}
\caption{The SONGS Unit 2 Reactor ramping from zero to full power over the course of several days.} \label{fig:power}
\end{minipage}\hspace{2pc}%
\begin{minipage}{3in}
\includegraphics*[width=2.5in]{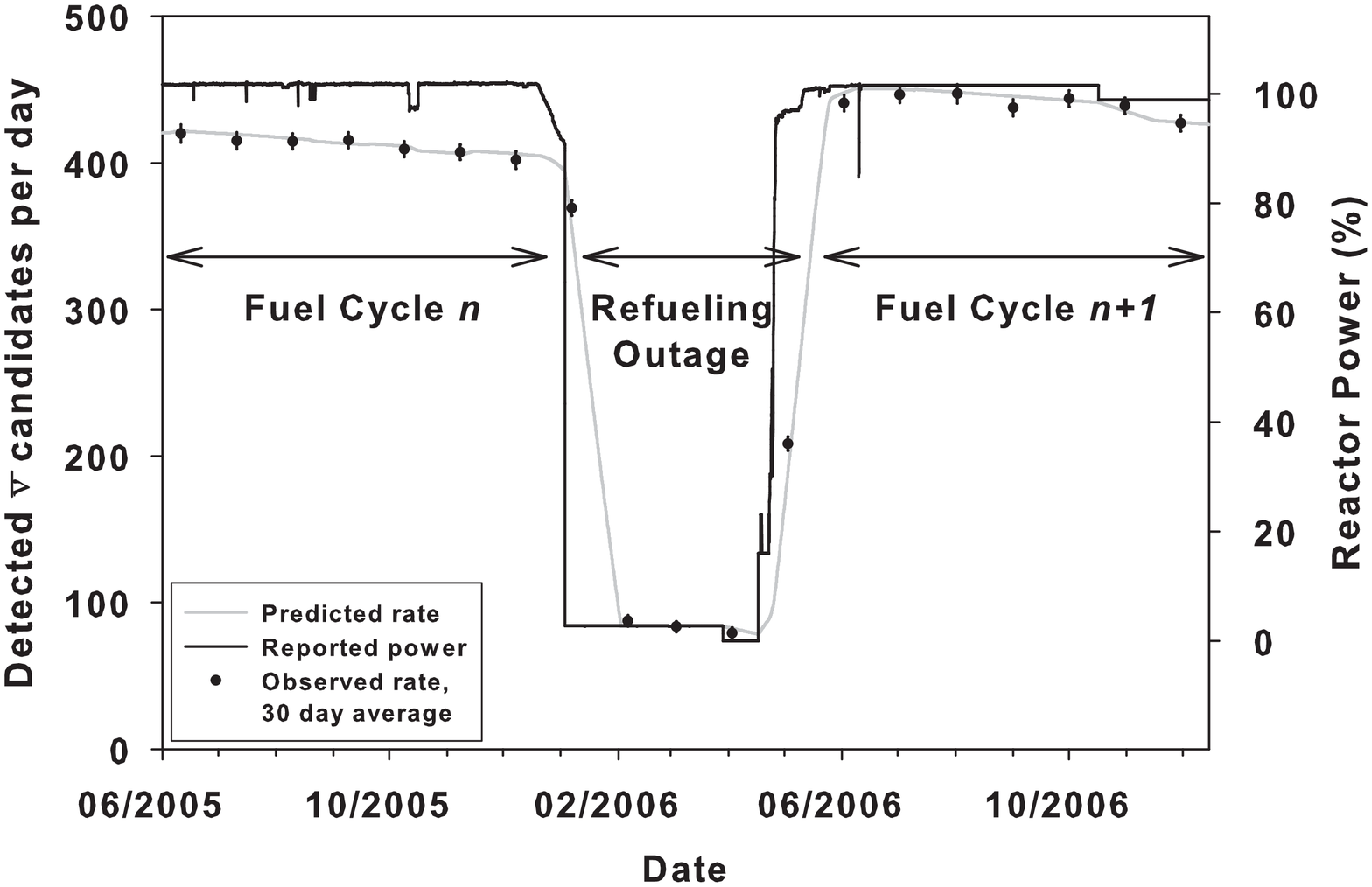}
\caption{Antineutrino rate measurements before, during, and after a rector refueling outage. The decrease in detection rate as the fuel evolves and the step increase in rate after refueling can be seen.} \label{fig:burnup}
\end{minipage}
\end{figure}

The SONGS1 detector was operated in a completely automatic fashion. Automatic calibration and analysis procedures were implemented and antineutrino detection rate data was transmitted to SNL/LLNL in near real time. An example of the ability to track changes in reactor thermal power is given in Fig.~\ref{fig:power}. A reactor scram (emergency shutdown) could be observed within 5 hours of its occurrence at $99.9\%$ confidence. Integrating the antineutrino detection rate data over a $24$~hour period yielded a relative power monitoring precision of about $8\%$, while increasing the averaging period to $7$~days yielded a precision of about $3\%$~\cite{powerpaper}.

Increasing the averaging time to $30$~days allowed observation of the fuel burnup~\cite{burnuppaper} (Fig.~\ref{fig:burnup}). The relatively simple calibration procedure was able to maintain constant detector efficiency to better than 1\% over the $18$~month observation period. The decrease in rate due to fuel evolution (burnup) and the step increase in rate expected after refueling (exchange of Pu laden fuel for fresh fuel containing only U) were both clearly observed.

\subsubsection{Non-flammable detectors}

SONGS1 was very successful, demonstrating that a relatively simple and compact detector could be operated non-intrusively at a commercial reactor for a long period. However, much of the feedback received from the safeguards community focussed upon the use of a flammable liquid scintillator. As used in the SONGS1 deployment this scintillator presented no safety hazard to the operation of the reactor - all relevant safety codes and procedures were checked and strictly adhered to. However, deployment of that flammable material did require some compliance effort from the reactor operator. In a safeguards context such situations should be avoided if at all possible.

Therefore, we decided to develop and deploy two detectors based upon non-flammable materials. The first of these was based upon a plastic scintillator material. Our goal was retain as much similarity between this device and SONGS1 as possible. Therefore we wished to use Gd as the neutron capture agent; this was achieved by using $24$ $1$~m~x~$0.75$~m~x~$2$~cm slabs of BC-408 plastic scintillator and interleaving them with a Gd loaded layer. The total active volume of this detector was $\approx 0.36$~m$^3$. This device was deployed at SONGS in 2007, and was clearly able to observe reactor outages like that shown in Fig.~\ref{fig:power}. We continue to analyze the data from this detector and expect to be able to observe fuel burnup also.

%The second non-flammable device is a water Cerenkov detector.
Inspired by the GADZOOKS concept presented at this conference in 2004~\cite{gadzooks}, we are also investigating the use of Gd-doped water as an antineutrino detector. This detection medium should have the advantage of being insensitive to the correlated background produced by cosmogenic fast neutrons that recoil from a proton and then capture. A $250$~liter tank of purified water containing 0.1\% Gd by weight was been deployed in the SONGS tendon gallery. Initially, this detector was deployed with little passive shielding - the large uncorrelated background rate that resulted has made it difficult to identify a reactor antineutrino signal much beyond a $3\sigma$ level.
%Data from a deployment period with passive shielding in place is still to be analyzed.
In this configuration Gd-doped water also appears promising for use in Special Nuclear Material search applications~\cite{water}.

\subsubsection{Observation and application of neutrino-nucleus coherent scattering}

%The inverse beta reaction is, in theory, not the only means by which reactor antineutrinos can be detected.
The coherent scattering of an (anti)neutrino from a nucleus is a long predicted but as yet unobserved standard model process. It is difficult to observe since the signal is a recoiling nucleus with just a few keV of energy. Nonetheless, this process holds great promise for reactor monitoring since it has  cross-section several orders of magnitude higher than that for inverse beta decay, which might eventually yield significantly smaller monitoring detectors. To explore the prospects for this process, we are currently collaborating with the Collar group of the University of Chicago~\cite{collar} in deploying an ultra-low threshold Ge crystal at SONGS. We are also investigating the potential of dual phase argon detectors~\cite{coherent}.

\subsection{Effort in France}

\subsubsection{Double Chooz}

The Double Chooz collaboration~\cite{DCLOI} plans to use the Double Chooz near detector ($\approx 400$~m from the two Chooz reactors) for a precision non-proliferation measurement~\cite{DC_near}. The Double Chooz detectors will represent the state-of-the-art in antineutrino detection, and will be able to make a benchmark measurement of the antineutrino energy spectrum emitted by a commercial PWR. There is also a significant effort underway within Double Chooz to improve the reactor simulations used to predict reactor fission rates and the measurements of the antineutrino energy spectrum emitted by the important fissioning isotopes. This work is necessary for the physics goals of Double Chooz, but it will also greatly improve the precision with which the fuel evolution of a reactor can be measured.

\subsubsection{Nucifer}

The Double Chooz near detector design is too complex and costly for widespread safeguards use. Therefore, the Double Chooz groups in CEA/Saclay, IN2P3-Subatech, and APC plan to apply the technology developed for Double Chooz, in particular detector simulation capabilities and high flash-point liquid scintillator, to the development of a compact antineutrino detector for safeguards: Nucifer~\cite{nucifer2}. The emphasis of this design will be on maintaining high detection efficiency ($\approx 55\%$) and good energy resolution and background rejection. Nucifer will be commissioned against research reactors in France during 2009-2010.
%at Saclay in 2009 (OSIRIS, 70~MW$_{th}$) and perhaps Grenoble (ILL) in 2010. This last location is particularly interesting as the fuel used in the ILL is 97\% $^{235}$U.
Following the commissioning phase Nucifer will be deployed against a commercial PWR, where it is planned to measure reactor fuel evolution using the antineutrino energy spectrum.

\subsection{Effort in Brazil}

An effort to develop a compact antineutrino detector for reactor safeguards is also underway in Brazil, at the Angra dos Rios Nuclear Power Plant~\cite{brazil1}. %This may be a precursor to a second generation theta-13 experiment.
Several deployment sites near the larger of the two reactors at Angra have been negotiated with the plant operator and detector design is well underway. This effort is particularly interesting, as a third reactor is soon to be built at the Angra site, within which space may be specially reserved for a detector, and because in addition to the IAEA, there is a regional safeguards organization (ABBAC) monitoring operations. Such regional agencies often pioneer the use of new safeguards technologies, and the detector deployment at Angra may occur with ABBAC involvement.

\subsection{Effort in Japan}

A prototype detector for the KASKA theta-13 experiment has been deployed at the Joyo fast research reactor in Japan~\cite{Japan}. This effort is notable, since it is an attempt to observe antineutrinos with a compact detector at a small research reactor in a deployment location with little overburden. This may be typical of the challenging environment in which the IAEA has recently expressed interest in applying this technique (Sec.~\ref{sec:iaea}). Not surprisingly, this effort has encountered large background rates, and to date has identified no clear reactor antineutrino signal.

\section{IAEA Interest}
\label{sec:iaea}

The IAEA is aware of the developments occurring in this field. A representative from the IAEA Novel Technologies group attended the most recent AAP meeting, and expressed an interest in using this monitoring technique at research reactors. The IAEA currently uses a device at these facilities that requires access to the primary coolant loop~\cite{IAEA}. Needless to say, this is quite invasive and an antineutrino monitoring technique would clearly be superior in this respect.
%The relatively small power of research reactors will make monitoring these facilities challenging, but closer access to the core may offset this to some extent. %
An experts meeting at IAEA headquarters will occur in October of 2008 to discuss the capabilities of current and projected antineutrino detection techniques and the needs of the IAEA.

\section{Conclusion}
\label{sec:conclusion}
\emph{Applications} of neutrino physics must have seemed somewhat fanciful when first discussed at this conference several decades ago. But even with currently available technologies, useful reactor monitoring appears feasible, as demonstrated by the SONGS1 results. The IAEA has expressed interest in this technique and the Applied Antineutrino Physics community eagerly awaits their guidance as to the steps required to add antineutrino based reactor monitoring to the safeguards toolbox.

LLNL-PROC-406953.

This work was performed under the auspices of the U.S. Department of Energy by Lawrence Livermore National Laboratory in part under Contract W-7405-Eng-48 and in part under Contract DE-AC52-07NA27344.

\end{document}